\documentstyle[12pt,prd,aps,epsfig,preprint]{revtex}

\def\bea{\begin{eqnarray}}
\def\eea{\end{eqnarray}}

\begin{document}

\begin{titlepage}

\title{The Effect of Supersymmetric CP Phases on $q \overline{q}$ Annihilations}
\author{Aytekin AYDEMIR}
\address{Dept. of Physics, Mersin University, Mersin- Turkey}

\author{Kerem CANKOCAK}
\address{Department of Physics, Mugla University, Mugla- Turkey}

\author{Ramazan SEVER}
\address{Dept. of Physics, Middle East Technical University, Ankara- Turkey}
\date{\today }

\thispagestyle{empty}

\baselineskip=15pt

\thispagestyle{empty}

\maketitle

\begin{abstract}
We compute the rates for $q\ \overline{q}$ annihilation into
charginos and neutralinos by taking into account the effects
of supersymmetric soft phases. In particular, the phase of
the $\mu$ parameter gains direct accessibility via the production
of dissimilar charginos and neutralinos. The phases of
the trilinear soft masses do not have a significant effect
on the cross sections. Our results can be important for sparticle
searches at the LHC.

\end{abstract}




\end{titlepage}
\newpage

\section{Introduction}

Supersymmetry (SUSY), is one of the most favored extensions of
the SM which is capable of stabilizing the ino-sector of
fundamental scalars against the ultraviolet divergences. The
(soft) breaking of SUSY, around the ${\rm TeV}$ scale, brings
about two new ingredients compared to the standard electroweak
theory (SM): First, there are novel sources of flavor violation
coming through the off--diagonal entries of the squark mass
matrices. Second, there are novel sources of CP violation coming
from the phases of the soft masses. The first effect, which cannot
be determined theoretically, is strongly constrained by the FCNC
data \cite{masiero} , and therefore, as a predictive case, it is
convenient to restrict all flavor--violating transitions to the
charged--current interactions where they proceed via the known CKM
angles. However, this very restriction of the flavor violation to
the SM one does not evade new sources of CP violation. Indeed, the
model possesses various flavor--blind CP--odd phases contained in
the complex $\mu $ parameter, $A$ parameters, and gauge fermion
masses $M_{i}$.

These phases form the new sources of CP violation which shows up
in the electric dipole moments (EDMs) of leptons and hadrons (See
\cite{edm1} and references therein). For heavy quark EDMs see
\cite{edm2}) and for the rate asymmetries of various heavy--light
mesons \cite{meson}. Therefore, it is of fundamental importance to
determine appropriate collider processes where all or some of the
SUSY CP phases can be inferred or measured. In fact, the effects
of the SUSY CP phases on the Higgs production have been already
analyzed in \cite{higgs,higgsp}. In this work we will discuss the
chargino and neutralino production at LHC energies and ways of
isolating the phase of the $\mu $ parameter from the cross
section. We shall compute the
cross section for $q\bar{q}\rightarrow \tilde{\chi}_{i}^{+}\tilde{\chi}%
_{j}^{-}$ as a function of $\varphi _{\mu }=\mbox{Arg}[\mu ]$ for
various values of $|\mu |$ and the SU(2) gaugino masses
$M_{1},M_{2}$. We shall also
compute the cross section for $q\bar{q}\rightarrow \tilde{\chi}_{i}^{0}%
\tilde{\chi}_{j}^{0}$ for various SUSY parameters.

\section{$q\bar{q}\rightarrow \tilde{\protect\chi}_{i}^{+}\tilde{\protect\chi%
}_{j}^{-}$}

Our analysis is similar to that used for the linear collider
processes \cite{lc}. The relevant Feynman diagrams are depicted in
Fig. 1. In what follows we mainly deal with the first two diagrams
since the third one is suppressed by presumably heavy squarks.
Then it is obvious that the amplitude for the process depends
exclusively on the phases in the chargino sector, $i.e$, the phase
of the $\mu $ parameter.

Here we summarize the masses and couplings of the charginos for completeness
(See \cite{phase} for details). The charginos which are the mass eigenstates
of charged gauginos and Higgsinos are described by a $2\times 2$ mass matrix
\begin{equation}
M_{C}=\left(
\begin{array}{cc}
M_{2} & \sqrt{2}M_{W}\cos \beta \\
\sqrt{2}M_{W}\sin \beta & |\mu |e^{i\varphi _{\mu }}
\end{array}
\right)
\end{equation}
where $M_{2}$ is the SU(2) gaugino mass taken to be real throughout the
work. The masses of the charginos as well as their mixing matrices follow
from the bi-unitary transformation
\begin{equation}
C_{R}^{\dagger }M_{C}C_{L}=\mbox{diag}(m_{\chi _{1}},m_{\chi _{2}})
\end{equation}
where $C_{L}$ and $C_{R}$ are $2\times 2$ unitary matrices, and $m_{\chi
_{1}}$, $m_{\chi _{2}}$ are the masses of the charginos $\chi _{1}$, $\chi
_{2}$ such that $m_{\chi _{1}}<m_{\chi _{2}}$. It is convenient to choose
the following explicit parametrization for the chargino mixing matrices:
\begin{eqnarray}
C_{L} &=&\left(
\begin{array}{cc}
\cos \theta _{L} & \sin \theta _{L}e^{i\varphi _{L}} \\
-\sin \theta _{L}e^{-i\varphi _{L}} & \cos \theta _{L}
\end{array}
\right) \\
C_{R} &=&\left(
\begin{array}{cc}
\cos \theta _{R} & \sin \theta _{R}e^{i\varphi _{R}} \\
-\sin \theta _{R}e^{-i\varphi _{R}} & \cos \theta _{R}
\end{array}
\right) \cdot \left(
\begin{array}{cc}
e^{i\phi _{1}} & 0 \\
0 & e^{i\phi _{2}}
\end{array}
\right)
\end{eqnarray}
where the angle parameters $\theta _{L,R}$, $\varphi _{L,R}$, and $\phi
_{1,2}$ can be determined from the defining equation (1). A straightforward
calculation yields
\begin{eqnarray}
\tan 2\theta _{L} &=&\frac{\sqrt{8}M_{W}\sqrt{M_{2}^{2}\cos ^{2}\beta +|\mu
|^{2}\sin ^{2}\beta +|\mu |M_{2}\sin 2\beta \cos \varphi _{\mu }}}{%
M_{2}^{2}-|\mu |^{2}-2M_{W}^{2}\cos 2\beta }  \nonumber \\
\tan 2\theta _{R} &=&\frac{\sqrt{8}M_{W}\sqrt{|\mu |^{2}\cos ^{2}\beta
+M_{2}^{2}\sin ^{2}\beta +|\mu |M_{2}\sin 2\beta \cos \varphi _{\mu }}}{%
M_{2}^{2}-|\mu |^{2}+2M_{W}^{2}\cos 2\beta }  \nonumber \\
\tan \varphi _{L} &=&\frac{|\mu |\sin \varphi _{\mu }}{M_{2}\cot \beta +|\mu
|\cos \varphi _{\mu }}  \nonumber \\
\tan \varphi _{R} &=&-\frac{|\mu |\cot \beta \sin \varphi _{\mu }}{|\mu
|\cot \beta \cos \varphi _{\mu }+M_{2}}
\end{eqnarray}
in terms of which the remaining two angles $\phi _{1}$ and $\phi _{2}$ read
as follows
\begin{equation}
\tan \phi _{i}=\frac{\mbox{Im}[Q_{i}]}{\mbox{Re}[Q_{i}]}
\end{equation}
where $i=1,2$ and
\begin{eqnarray}
Q_{1} &=&\sqrt{2}M_{W}[\cos \beta \sin \theta _{L}\cos \theta
_{R}e^{-i\varphi _{L}}+\sin \beta \cos \theta _{L}\sin \theta
_{R}e^{i\varphi _{R}}]  \nonumber \\
&+&M_{2}\cos \theta _{L}\cos \theta _{R}+|\mu |\sin \theta _{L}\sin \theta
_{R}e^{i(\varphi _{\mu }+\varphi _{R}-\varphi _{L})}  \nonumber \\
Q_{2} &=&-\sqrt{2}M_{W}[\cos \beta \sin \theta _{R}\cos \theta
_{L}e^{-i\varphi _{R}}+\sin \beta \cos \theta _{R}\sin \theta
_{L}e^{i\varphi _{L}}]  \nonumber \\
&+&M_{2}\sin \theta _{L}\sin \theta _{R}e^{i(\varphi _{L}-\varphi
_{R})}+|\mu |\cos \theta _{L}\cos \theta _{R}e^{i\varphi _{\mu }}\,.
\end{eqnarray}
The origin of the phases $\theta _{L,R}$, $\varphi _{L,R}$, and $\phi _{1,2}$
is easy to trace back. The angles $\theta _{L}$ and $\theta _{R}$ would be
sufficient to diagonalize, respectively, the quadratic mass matrices $%
M_{C}^{\dagger }M_{C}$ and $M_{C}M_{C}^{\dagger }$ if $M_{C}$ were real. As
a result one needs the additional phases $\varphi _{L,R}$ which are
identical to the phases in the off--diagonal entries of the matrices $%
M_{C}^{\dagger }M_{C}$ and $M_{C}M_{C}^{\dagger }$, respectively. However,
these four phases are still not sufficient for making the chargino masses
real positive due to the bi-unitary nature of the transformation, and hence,
the phases $\phi _{1}$ and $\phi _{2}$ can not also be made real positive.
Finally, inserting the unitary matrices $C_{L}$ and $C_{R}$ into the
defining equation (1) one obtains the following expressions for the masses
of the charginos
\begin{eqnarray}
m_{\chi _{1(2)}}^{2} &=&\frac{1}{2}\Big\{M_{2}^{2}+|\mu
|^{2}+2M_{W}^{2}-(+)[(M_{2}^{2}-|\mu |^{2})^{2}+4M_{W}^{4}\cos
^{2}2\beta
\nonumber \\
&+&4M_{W}^{2}(M_{2}^{2}+|\mu |^{2}+2M_{2}|\mu |\sin 2\beta \cos \varphi
_{\mu })]^{1/2}\Big\}.
\end{eqnarray}

The fundamental SUSY parameters $M_{2}$, $|\mu |$, $\tan \beta $ and the
phase parameter $\cos \varphi _{\mu }$ can be extracted from the chargino $%
\tilde{\chi}_{1,2}^{\pm }$ parameters \cite{lc} $i.e.$ the masses $m_{\tilde{%
\chi}_{1,2}^{\pm }}$ and the two mixing angles $\phi _{L}$ and $\phi _{R}$
of the left and right chiral components of the wave function. These mixing
angles are physical observables and they can be measured just like the
chargino masses $m_{\tilde{\chi}_{1,2}^{\pm }}$ in the process $q+\bar{q}%
\rightarrow \tilde{\chi}_{i}^{+}+\tilde{\chi}_{j}^{-}$. The two angles $\phi
_{L}$ and $\phi _{R}$ and the nontrivial phase angles $\{\varphi _{L},$ $%
\varphi _{R},$ $\phi _{1},\phi _{2}\}$ define the couplings of the
chargino-chargino-Z vertices:

\[
\left\langle \tilde{\chi}_{1L}^{-}\left| Z\right| \tilde{\chi}%
_{1L}^{-}\right\rangle =-\frac{e}{s_{W}c_{W}}[s_{W}^{2}-\frac{3}{4}-\frac{1}{%
4}\cos 2\theta _{L}]
\]

\[
\left\langle \tilde{\chi}_{1L}^{-}\left| Z\right| \tilde{\chi}%
_{2L}^{-}\right\rangle =+\frac{e}{4s_{W}c_{W}}e^{-i\varphi _{L}}\sin 2\theta
_{L}
\]

\[
\left\langle \tilde{\chi}_{2L}^{-}\left| Z\right| \tilde{\chi}%
_{2L}^{-}\right\rangle =-\frac{e}{s_{W}c_{W}}[s_{W}^{2}-\frac{3}{4}+\frac{1}{%
4}\cos 2\theta _{L}]
\]

\[
\left\langle \tilde{\chi}_{1R}^{-}\left| Z\right| \tilde{\chi}%
_{1R}^{-}\right\rangle =-\frac{e}{s_{W}c_{W}}[s_{W}^{2}-\frac{3}{4}-\frac{1}{%
4}\cos 2\theta _{R}]
\]

\[
\left\langle \tilde{\chi}_{1R}^{-}\left| Z\right| \tilde{\chi}%
_{2R}^{-}\right\rangle =+\frac{e}{4s_{W}c_{W}}e^{-i(\varphi _{R}-\phi
_{1}+\phi _{2})}\sin 2\theta _{R}
\]

\begin{equation}
\left\langle \tilde{\chi}_{2R}^{-}\left| Z\right| \tilde{\chi}%
_{2R}^{-}\right\rangle =-\frac{e}{s_{W}c_{W}}[s_{W}^{2}-\frac{3}{4}+\frac{1}{%
4}\cos 2\theta _{R}]
\end{equation}
where $s_{W}=\sin \theta _{W}$ is the weak angle. Note that every vertex
here is an explicit function of $\varphi _{\mu }$ via the various mixing
angles. However, the $Z$ coupling to unlike charginos $\tilde{\chi}_{i}^{+}%
\tilde{\chi}_{j}^{-}$ is manifestly complex, and its phase vanishes in the
CP--conserving limit, $\varphi _{\mu }\rightarrow 0,$ $\pi $.

Obviously, the photon vertex is independent of the SUSY phases:
\begin{equation}
\left\langle \tilde{\chi}_{iL,R}^{-}\left| \gamma \right| \tilde{\chi}%
_{jL,R}^{-}\right\rangle =e\delta _{ij}
\end{equation}

The process $q\bar{q}\rightarrow
\tilde{\chi}_{i}^{+}\tilde{\chi}_{j}^{-}$ is generated by the two
mechanisms shown in Fig. 1.: the $s$-channel $\gamma $ and $Z$
exchanges, and $t$-channel $\tilde{q}$ exchange, where the latter
is consistently neglected below. The transition amplitude can be
parameterized as
\begin{equation}
T(q\bar{q}\rightarrow \tilde{\chi}_{i}^{+}\tilde{\chi}_{j}^{-})=\frac{e^{2}}{%
s}Q_{\alpha \beta }[\bar{v}(\bar{q})\gamma _{\mu }P_{\alpha }u(q)][\bar{u}(%
\tilde{\chi}_{i}^{-})\gamma ^{\mu }P_{\beta }v(\tilde{\chi}_{j}^{+})]
\end{equation}
where the charges $Q_{\alpha \beta }$ are defined such that the first index
corresponds to the chirality of the $\overline{q}q$ current and the second
one to chargino current. For various final states, their expressions are
given by:

(i)\qquad \underline{$\tilde{\chi}_{1}^{-}\tilde{\chi}_{1}^{+}$} for
\underline{$q=u,c$}

\[
Q_{LL}=1+\frac{D_{Z}}{s_{W}^{2}c_{W}^{2}}(\frac{1}{2}-\frac{2}{3}%
s_{W}^{2})(s_{W}^{2}-\frac{3}{4}-\frac{1}{4}\cos 2\phi _{L})
\]

\[
Q_{LR}=1+\frac{D_{Z}}{s_{W}^{2}c_{W}^{2}}(\frac{1}{2}-\frac{2}{3}%
s_{W}^{2})(s_{W}^{2}-\frac{3}{4}-\frac{1}{4}\cos 2\phi _{R})
\]

\[
Q_{RL}=1+\frac{D_{Z}}{c_{W}^{2}}(-\frac{2}{3})(s_{W}^{2}-\frac{3}{4}-\frac{1%
}{4}\cos 2\phi _{L})
\]

\begin{equation}
Q_{RR}=1+\frac{D_{Z}}{c_{W}^{2}}(-\frac{2}{3})(s_{W}^{2}-\frac{3}{4}-\frac{1%
}{4}\cos 2\phi _{R})
\end{equation}

(ii)\qquad \underline{$\tilde{\chi}_{1}^{-}\tilde{\chi}_{1}^{+}$} for
\underline{$q=d,s$}

\[
Q_{LL}=1+\frac{D_{Z}}{s_{W}^{2}c_{W}^{2}}(-\frac{1}{2}+\frac{1}{3}%
s_{W}^{2})(s_{W}^{2}-\frac{3}{4}-\frac{1}{4}\cos 2\phi _{L})
\]

\[
Q_{LR}=1+\frac{D_{Z}}{s_{W}^{2}c_{W}^{2}}(-\frac{1}{2}+\frac{1}{3}%
s_{W}^{2})(s_{W}^{2}-\frac{3}{4}-\frac{1}{4}\cos 2\phi _{R})
\]

\[
Q_{RL}=1+\frac{D_{Z}}{c_{W}^{2}}(+\frac{1}{3})(s_{W}^{2}-\frac{3}{4}-\frac{1%
}{4}\cos 2\phi _{L})
\]

\begin{equation}
Q_{RR}=1+\frac{D_{Z}}{c_{W}^{2}}(+\frac{1}{3})(s_{W}^{2}-\frac{3}{4}-\frac{1%
}{4}\cos 2\phi _{R})
\end{equation}
\bigskip

(iii)\qquad \underline{$\tilde{\chi}_{1}^{-}\tilde{\chi}_{2}^{+}$} for
\underline{$q=u,c$}

\[
Q_{LL}=\frac{D_{Z}}{4s_{W}^{2}c_{W}^{2}}(\frac{1}{2}-\frac{2}{3}%
s_{W}^{2})e^{-i\varphi _{L}}\sin 2\phi _{L}
\]

\[
Q_{LR}=\frac{D_{Z}}{4s_{W}^{2}c_{W}^{2}}(\frac{1}{2}-\frac{2}{3}%
s_{W}^{2})e^{-i(\varphi _{R}-\phi _{1}+\phi _{2})}\sin 2\phi _{R}
\]

\[
Q_{RL}=\frac{D_{Z}}{4c_{W}^{2}}(-\frac{2}{3})e^{-i\varphi _{L}}\sin 2\phi
_{L}
\]

\begin{equation}
Q_{RR}=\frac{D_{Z}}{4c_{W}^{2}}(-\frac{2}{3})e^{-i(\varphi _{R}-\phi
_{1}+\phi _{2})}\sin 2\phi _{R}
\end{equation}

(iv)\qquad \underline{$\tilde{\chi}_{1}^{-}\tilde{\chi}_{2}^{+}$} for
\underline{$q=d,s$}

\[
Q_{LL}=\frac{D_{Z}}{4s_{W}^{2}c_{W}^{2}}(-\frac{1}{2}+\frac{1}{3}%
s_{W}^{2})e^{-i\varphi _{L}}\sin 2\phi _{L}
\]

\[
Q_{LR}=\frac{D_{Z}}{4s_{W}^{2}c_{W}^{2}}(-\frac{1}{2}+\frac{1}{3}%
s_{W}^{2})e^{-i(\varphi _{R}-\phi _{1}+\phi _{2})}\sin 2\phi _{R}
\]

\[
Q_{RL}=\frac{D_{Z}}{4c_{W}^{2}}(+\frac{1}{3})e^{-i\varphi _{L}}\sin 2\phi
_{L}
\]

\begin{equation}
Q_{RR}=\frac{D_{Z}}{4c_{W}^{2}}(+\frac{1}{3})e^{-i(\varphi _{R}-\phi
_{1}+\phi _{2})}\sin 2\phi _{R}
\end{equation}

(v)\qquad \underline{$\tilde{\chi}_{2}^{-}\tilde{\chi}_{2}^{+}$} for
\underline{$q=u,c$}

\[
Q_{LL}=1+\frac{D_{Z}}{s_{W}^{2}c_{W}^{2}}(\frac{1}{2}-\frac{2}{3}%
s_{W}^{2})(s_{W}^{2}-\frac{3}{4}+\frac{1}{4}\cos 2\phi _{L})
\]

\[
Q_{LR}=1+\frac{D_{Z}}{s_{W}^{2}c_{W}^{2}}(\frac{1}{2}-\frac{2}{3}%
s_{W}^{2})(s_{W}^{2}-\frac{3}{4}+\frac{1}{4}\cos 2\phi _{R})
\]

\[
Q_{RL}=1+\frac{D_{Z}}{c_{W}^{2}}(-\frac{2}{3})(s_{W}^{2}-\frac{3}{4}+\frac{1%
}{4}\cos 2\phi _{L})
\]

\begin{equation}
Q_{RR}=1+\frac{D_{Z}}{c_{W}^{2}}(-\frac{2}{3})(s_{W}^{2}-\frac{3}{4}+\frac{1%
}{4}\cos 2\phi _{R})
\end{equation}

(vi)\qquad \underline{$\tilde{\chi}_{2}^{-}\tilde{\chi}_{2}^{+}$} for
\underline{$q=d,s$}

\[
Q_{LL}=1+\frac{D_{Z}}{s_{W}^{2}c_{W}^{2}}(-\frac{1}{2}+\frac{1}{3}%
s_{W}^{2})(s_{W}^{2}-\frac{3}{4}+\frac{1}{4}\cos 2\phi _{L})
\]

\[
Q_{LR}=1+\frac{D_{Z}}{s_{W}^{2}c_{W}^{2}}(-\frac{1}{2}+\frac{1}{3}%
s_{W}^{2})(s_{W}^{2}-\frac{3}{4}+\frac{1}{4}\cos 2\phi _{R})
\]

\[
Q_{RL}=1+\frac{D_{Z}}{c_{W}^{2}}(+\frac{1}{3})(s_{W}^{2}-\frac{3}{4}+\frac{1%
}{4}\cos 2\phi _{L})
\]

\begin{equation}
Q_{RR}=1+\frac{D_{Z}}{c_{W}^{2}}(+\frac{1}{3})(s_{W}^{2}-\frac{3}{4}+\frac{1%
}{4}\cos 2\phi _{R})
\end{equation}
Here all the amplitudes are built up by the $\gamma$ and $Z$ exchanges, and $%
D(Z)$ stands for the $Z$ propagator: $D_{Z}=s/(s-m_{Z}^{2}+im_{Z}\Gamma
_{Z}) $.

In what follows, for convenience we will introduce four combinations of the
charges
\begin{eqnarray}
Q_{1} &=&\frac{1}{4}[\left| Q_{RR}\right| ^{2}+\left| Q_{LL}\right|
^{2}+\left| Q_{RL}\right| ^{2}+\left| Q_{LR}\right| ^{2}]  \nonumber \\
Q_{2} &=&\frac{1}{2}Re[Q_{RR}Q_{RL}^{\ast }+Q_{LL}Q_{LR}^{\ast }]  \nonumber
\\
Q_{3} &=&\frac{1}{4}[\left| Q_{RR}\right| ^{2}+\left| Q_{LL}\right|
^{2}-\left| Q_{RL}\right| ^{2}-\left| Q_{LR}\right| ^{2}]  \nonumber \\
Q_{4} &=&\frac{1}{2}Im[Q_{RR}Q_{RL}^{\ast }+Q_{LL}Q_{LR}^{\ast }]
\end{eqnarray}
so that the differential cross section can be expressed simply as
\begin{equation}
\frac{d\sigma }{d\cos \Theta }(q\bar{q}\rightarrow \tilde{\chi}_{i}^{+}%
\tilde{\chi}_{j}^{-})=\frac{\pi \alpha ^{2}}{2s}\lambda ^{1/2}\{[1-(\mu
_{i}^{2}-\mu _{j}^{2})^{2}+\lambda \cos ^{2}\Theta ]Q_{1}+4\mu _{i}\mu
_{j}Q_{2}+2\lambda ^{1/2}Q_{3}\cos \Theta \}
\end{equation}
with the usual two body phase space factor:
\begin{equation}
\lambda (1,\mu _{i}^{2},\mu _{j}^{2})=[1-(\mu _{i}+\mu _{j})^{2}][1-(\mu
_{i}-\mu _{j})^{2}]
\end{equation}
defined via the reduced mass $\mu
_{i}^{2}=m_{\tilde{\chi}_{i}^{\pm }}^{2}/{s}$.

Integrating the differential cross section over the center--of--mass
scattering angle $\Theta $ we arrive at the total cross section
\begin{equation}
\sigma =\sigma (\varphi _{\mu },\mu ,M_{2},s,\tan \beta )
\end{equation}
whose dependencies on $\varphi _{\mu },M_{2}$ and $|\mu |$ will be
analyzed numerically.

Besides the total cross section, it is necessary to analyze the
rate asymmetries for having better information about $\varphi
_{\mu }$. Concerning this point, we investigate the normal
polarization vector of the charginos which are inherently CP--odd
and exist therefore if CP is broken in the fundamental theory.

Defining the polarization vector $\vec{P}=({P}_L,{P}_T, {P}_N)$ in
the rest frame of the chargino, where ${P}_L$ denotes the
component parallel to the charginos flight direction, ${P}_T$ the
transverse component in the production plane, and ${P}_N$ is the
component normal to the production plane, these three components
can be expressed by helicity amplitudes in the following
way\cite{llc}:
\begin{eqnarray}
&& {P}_L=\frac{1}{4}\sum_{\sigma=\pm}\left\{
              |\langle\sigma;++\rangle|^2+|\langle\sigma;+-\rangle|^2
             -|\langle\sigma;-+\rangle|^2-|\langle\sigma;--\rangle|^2
                                           \right\}/{N}
              \nonumber\\
&& {P}_T=\frac{1}{2}{\rm Re}\bigg\{\sum_{\sigma=\pm}\left[
              |\langle\sigma;++\rangle\langle\sigma;-+\rangle^*
             +|\langle\sigma;--\rangle\langle\sigma;+-\rangle^*
                          \right]\bigg\}/{N}\nonumber\\
&& {P}_N=\frac{1}{2}{\rm Im}\bigg\{\sum_{\sigma=\pm}\left[
              |\langle\sigma;--\rangle\langle\sigma;+-\rangle^*
             -|\langle\sigma;++\rangle\langle\sigma;-+\rangle^*
                           \right]\bigg\}/{N}
\end{eqnarray}
The longitudinal and transverse components are P--odd and
CP--even. The normal component is P--even and CP--odd, and it
 can  be generated by complex production amplitudes, c.f. Ref.\cite{Pnorm}.

Therefore, the normal polarization vector is defined as:
\begin{equation}
P_{N}=8\lambda ^{1/2}\mu _{j}\sin \Theta \frac{Q_{4}}{N}
\end{equation}
for $\tilde{\chi}_{j}^{+}\tilde{\chi}_{j}^{-},$ the $j$--th chargino, and is
defined as:

\begin{eqnarray}
P_{N}[\tilde{\chi}_{i,j}^{\pm }] &=&\pm 4\lambda ^{1/2}\mu
_{j,i}(F_{R}^{2}-F_{L}^{2})\sin \Theta \sin 2\phi _{L}  \nonumber
\\ &&\times \sin 2\phi _{R}\sin (\beta _{L}-\beta _{R}+\gamma
_{1}-\gamma _{2})/N
\end{eqnarray}
for non-diagonal pairs $\tilde{\chi}_{i}^{+}\tilde{\chi}_{j}^{-}$ where $%
i\neq j$. Here

\begin{equation}
N=4[(1-(\mu _{i}^{2}-\mu _{j}^{2})^{2}+\lambda \cos ^{2}\Theta
)Q_{1}+4\mu _{i}\mu _{j}Q_{2}+2\lambda ^{1/2}Q_{3}\cos \Theta ]
\end{equation}

and

\begin{equation}
F_{R}=\frac{D_{Z}}{4c_{W}^{2}},\text{ }F_{L}=\frac{D_{Z}}{4s_{W}^{2}c_{W}^{2}%
}(s_{W}^{2}-\frac{1}{2}).
\end{equation}

The polarization  can be measured from the two final--state
leptons, in the $\tilde{\chi}^\pm_{1,2}$ leptonic decays. A
non-vanishing $P_{N}$ will be sufficient to establish
non-vanishing CP violation in the system. Therefore, the value of
non-vanishing $P_{N}$ implies the strength of the CP invariance
braking in SUSY.

\section{\protect\bigskip $q\bar{q}\rightarrow \tilde{\protect\chi}_{i}^{0}%
\tilde{\protect\chi}_{j}^{0}$}

We also calculate neutralino pair production to investigate SUSY parameters,
since neutralino masses are relatively small to be produced at LHC energies.

The neutral supersymmetric fermionic partners of the B and W$^{3}$ gauge
bosons, $\widetilde{B}$ and $\widetilde{W}^{3},$ can mix with the neutral
supersymmetric fermionic partners of the Higgs bosons, $\widetilde{H_{1}}%
^{0} $ and $\widetilde{H_{2}}^{0}$ to form the mass eigenstates.

The neutralino mass matrix is

\begin{equation}
M_{N}=\left(
\begin{array}{c}
M_{1}e^{i\varphi _{1}} \\
0 \\
-m_{Z}s_{w}c_{\beta } \\
m_{Z}s_{w}s_{\beta }
\end{array}
\begin{array}{c}
0 \\
M_{2} \\
m_{Z}c_{w}c_{\beta } \\
-m_{Z}c_{w}s_{\beta }
\end{array}
\begin{array}{c}
-m_{Z}s_{w}c_{\beta } \\
m_{Z}c_{w}c_{\beta } \\
0 \\
\left| \mu \right| e^{i\varphi _{\mu }}
\end{array}
\begin{array}{c}
m_{Z}s_{w}s_{\beta } \\
-m_{Z}c_{w}s_{\beta } \\
\left| \mu \right| e^{i\varphi _{\mu }} \\
0
\end{array}
\right)
\end{equation}

whose diagonalization gives the physical states $\widetilde{\chi }_{i}^{0}$,
which are called neutralinos.

Since M$_{N}$ is a complex, symmetric matrix, it can be diagonalized by just
one unitary matrix N, such that

\begin{equation}
N^{\ast }M_{N}N^{\dagger }=diag(m_{\widetilde{\chi }_{1}^{0}},m_{\widetilde{%
\chi }_{2}^{0}},m_{\widetilde{\chi }_{3}^{0}},m_{\widetilde{\chi }_{4}^{0}}).
\end{equation}

The only Feynman diagram considered here is the last one given in Fig. 1,
that is s-channel Z exchange.

The matrix element for the process $q\bar{q}\rightarrow \tilde{\chi}_{i}^{0}%
\tilde{\chi}_{j}^{0}$ is

\begin{equation}
T(q\bar{q}\rightarrow \tilde{\chi}_{i}^{0}\tilde{\chi}_{j}^{0})=\frac{e^{2}}{%
s}Q_{\alpha \beta }^{ij}[\overline{v}(\overline{q})\gamma _{\mu }P_{\alpha
}u(q)]\times \lbrack \overline{u}(\tilde{\chi}_{i}^{0})\gamma ^{\mu
}P_{\beta }v(\tilde{\chi}_{j}^{0})]
\end{equation}

where the associated quark and neutralino currents are

\bigskip (i)\qquad \underline{$\tilde{\chi}_{1}^{0}\tilde{\chi}_{1}^{0}$}
for \underline{$q=u,c$}

\[
Q_{LL}=+\frac{D_{Z}}{s_{W}^{2}c_{W}^{2}}(\frac{1}{2}-\frac{2}{3}%
s_{W}^{2})Z_{11}
\]

\[
Q_{LR}=+\frac{D_{Z}}{s_{W}^{2}c_{W}^{2}}(\frac{1}{2}-\frac{2}{3}%
s_{W}^{2})Z_{11}^{\ast }
\]

\[
Q_{RL}=+\frac{D_{Z}}{c_{W}^{2}}(-\frac{2}{3})Z_{11}
\]

\begin{equation}
Q_{RR}=+\frac{D_{Z}}{c_{W}^{2}}(-\frac{2}{3})Z_{11}^{\ast }
\end{equation}

(ii)\qquad \underline{$\tilde{\chi}_{1}^{0}\tilde{\chi}_{1}^{0}$} for
\underline{$q=d,s$}

\[
Q_{LL}=+\frac{D_{Z}}{s_{W}^{2}c_{W}^{2}}(-\frac{1}{2}+\frac{1}{3}%
s_{W}^{2})Z_{11}
\]

\[
Q_{LR}=+\frac{D_{Z}}{s_{W}^{2}c_{W}^{2}}(-\frac{1}{2}+\frac{1}{3}%
s_{W}^{2})Z_{11}^{\ast }
\]

\[
Q_{RL}=+\frac{D_{Z}}{c_{W}^{2}}(+\frac{1}{3})Z_{11}
\]

\begin{equation}
Q_{RR}=+\frac{D_{Z}}{c_{W}^{2}}(+\frac{1}{3})Z_{11}^{\ast }
\end{equation}

and

\begin{equation}
\qquad Z_{ij}=\frac{1}{2}(N_{i3}N_{j3}^{\ast }-N_{i4}N_{j4}^{\ast })
\end{equation}
\qquad

\[
Z_{ij}^{\ast }=Z_{ij}
\]

Using the predefined charge combinations, differential cross section can be
expressed as

\begin{equation}
\frac{d\sigma }{d\cos \Theta }(q\bar{q}\rightarrow \tilde{\chi}_{1}^{0}%
\tilde{\chi}_{1}^{0})=\frac{\pi \alpha ^{2}}{2s}\lambda ^{1/2}\{[1+\lambda
\cos ^{2}\Theta ]Q_{1}+4\mu _{1}^{2}Q_{2}+2\lambda ^{1/2}Q_{3}\cos \Theta \}
\end{equation}

where

\begin{equation}
\lambda (1,\mu _{1}^{2},\mu _{1}^{2})=[1-4\mu _{1}{}^{2}]
\end{equation}
is defined via the reduced mass $\mu
_{1}^{2}=m_{\tilde{\chi}_{1}^{0}}^{2}/{s}$. $Q_{1},$ $Q_{2},$ and
$Q_{3}$ are as defined before.

Integrating the differential cross section over the
center--of--mass scattering angle $\Theta $ we arrive at the total
cross section
\begin{equation}
\sigma =\sigma (M_{1},\mu ,M_{2},s,\tan \beta )
\end{equation}
whose dependence on $M_{1}$, $|\mu |$ and s will be analyzed
numerically for $M_{2}=150$ GeV and $\tan \beta=30$.

\section{Numerical Estimates}

\subsection{Chargino production}

In this section we will discuss the dependence of the chargino
production cross section on $\varphi _{\mu },M_{2}$, $|\mu |$ and
$\sqrt{s}$. We everywhere apply the existing collider constraint
that $m_{\chi _{2}}>104\ GeV$.

In Table 1. we give the cross section values for $q\bar{q}\rightarrow \tilde{%
\chi}_{1}^{+}\tilde{\chi}_{1}^{-}$ and $q\bar{q}\rightarrow \tilde{\chi}%
_{1}^{+}\tilde{\chi}_{2}^{-}$ where M$_{2}=150,$ $200,$ $250$\ GeV, $\mu
=150,200,250$ GeV, $\tan \beta =4,$ $10,$ $30,$ $50,$ and $\varphi _{\mu
}=\pi /3$ are used in the calculations.

In Fig. 2 and Fig. 5 we show the dependence of the cross sections
$q\bar{q}\rightarrow \tilde{\chi}_{1}^{+}\tilde{\chi}_{1}^{-}$ and
$q\bar{q}\rightarrow \tilde{\chi}_{1}^{+}\tilde{\chi}_{2}^{-}$ on
$\varphi _{\mu }$\ for M$_{2}=150, 300$\ GeV, $\mu =150, 300$ GeV,
and $\tan \beta =$ 30.

For the process
$q\bar{q}\rightarrow\tilde{\chi}_{1}^{+}\tilde{\chi}_{1}^{-}$ ,
the dependence of the cross section on $\varphi _{\mu }$ is very
weak for the light charginos.  As seen from the Figs. 3 and 4, the
more spectacular enhancement implies the heavier chargino mass.

In  Fig. 6  we show the dependence of the cross sections
$q\bar{q}\rightarrow \tilde{\chi}_{1}^{+}\tilde{\chi}_{2}^{-}$ on
$\varphi _{\mu }$\ for $\mu =150$ GeV, and $\tan \beta =$ 30, as a
function of M$_{2}$. Again, there is small dependence of the cross
section on $\varphi _{\mu }$.

The increase of the cross section is tied up to the variation of
the chargino masses with the phases. It is clear that as $\varphi
_{\mu }:0\rightarrow \pi $ the mass splitting of the charginos
decrease, as expected from the Equation 8. This is an important
effect which implies that the cross section is larger than what
one would expect from the CP--conserving theory \cite{lc}.

Apart from the cross section itself, one can analyze various spin
and charge asymmetries which are expected to have an enhanced
dependence on $\varphi _{\mu }$. The normal polarization in
$q\bar{q}\rightarrow \tilde{\chi}_{1}^{+}\tilde{\chi}_{1}^{-}$ is
zero since the $\tilde{\chi}_{1}^{+}\tilde{\chi}_{1}^{-}\gamma $
and $\tilde{\chi}_{1}^{+}\tilde{\chi}_{1}^{-}Z$ vertices are real
even for non-zero phases in the chargino mass matrix.

In Fig. 7 we show the normal polarization $P_{N}$ of the unlike
charginos in $q\bar{q}\rightarrow
\tilde{\chi}_{1}^{+}\tilde{\chi}_{2}^{-}$\ which has a different
dependence on the phases. Here again M$_{2}=150,300$\ GeV, $ \mu
=150,300$ GeV, $\tan \beta =30,$ and $\varphi _{\mu }=\pi/2.$

The dependence of the normal polarization on the value of $\Theta$
and $\varphi _{\mu } $ \ is shown in Fig. 8, where the normal
polarization has its maximum at $\Theta = \pi/2$  as expected from
the Eqn. 23,  and at $\varphi _{\mu }=\pi/2$ as stated above.

However, we believe that for clarifying the essence of measuring $\varphi
_{\mu }$ the first quantity to be tested is the cross section itself.

\subsection{Neutralino production}

In this section, the dependence of the neutralino production
cross-section on center of mass energy $\sqrt{s}$ is investigated
for $\mu =150, 200$ GeV, $M_{1} =150, 200$ GeV and for $M_{2}
=150, 200$ GeV. In Fig. 9 we show the dependence of the cross
section for $q\bar{q}\rightarrow
\tilde{\chi}_{1}^{0}\tilde{\chi}_{1}^{0}$ on $\sqrt{s}$, for
$\varphi _{\mu }=0,\frac{\pi}{2}, \pi$  \ and $\frac{3\pi}{2}$,
where $M_{1} =150$ GeV, $M_{2} =100$ GeV and $\mu =200$ GeV. The
cross-section is maximum at about $\sqrt{s}=230$ GeV and drops as
$\sqrt{s}$ becomes higher. The variation of the cross section on
$\varphi _{\mu }$ is seen clearly. The highest cross section is
obtained for $\varphi _{\mu }=\pi$, when $M_{1}$ $<$ $\mu$, as
expected from the CP--conserving theory \cite{lc}.

 In Fig. 10, we show the dependence of the cross
section for $q\bar{q}\rightarrow
\tilde{\chi}_{1}^{0}\tilde{\chi}_{1}^{0}$ on $\sqrt{s}$ \ and \
$\mu$, for $\varphi _{\mu }=\frac{\pi}{2}$, $M_{1} =150$ GeV,
$M_{2} =150$ GeV and $\tan \beta =30 $.

 Finally, in Fig. 11, we plot the cross
section for $q\bar{q}\rightarrow
\tilde{\chi}_{1}^{0}\tilde{\chi}_{1}^{0}$ as a function of $\mu$ \
and \  $\varphi _{\mu }$, for $M_{1} =150$ GeV, $M_{2} =150$ GeV
and $\tan \beta =30 $. The dependence of the cross section on
$\varphi _{\mu }$ is seen clearly in this figure. At
$\varphi_{\mu}=\pi $ the cross section is lowest when $M_{1}$ $>$
$\mu$ and highest when $M_{1}$ $<$ $\mu$ respectively.

\section{Discussion and Conclusion}

We have analyzed the production of charginos and neutralinos at
LHC energies with the aim of isolating the phase of the $\mu $
parameter. The measurement of these processes will be an important
step for determining the CP violation sources of low--energy
supersymmetry. Our numerical results suggest that there is a
strong dependence on the phase of the $\mu$ parameter especially
when $|\mu|$ is comparable to the gaugino masses.

In true experimental environment, the cross sections we have
discussed above form the subprocess cross sections to be
integrated over appropriate structure functions. However, given
the energy span of LHC that it will be possible to probe
sparticles up to $2\ {\rm TeV}$, it is clear that the
center--of--mass energies we discuss are always within
experimental reach. If the experiment concludes $\varphi_{\mu}\sim
{\cal{O}}(1)$ then, given strong bounds from the absence of
permanent EDMs for electron, neutron, atoms and molecules, one
would conclude that the first two generations of sfermions will be
hierarchically split from the ones in the third generation. In
case the experiment reports a small $\varphi_{\mu}$ then
presumably all sfermion generations can lie right at the weak
scale in agreement with the EDM bounds. In this case, where
$\varphi_{\mu}$ is a small fraction of $\pi$, one might expect
that the minimal model is UV--completed above the TeV scale such
that the $\mu$ parameter is promoted to a dynamical SM--singlet
field ($e.g.$ the $Z^{\prime}$ models).

\newpage

FIGURE CAPTIONS

FIGURE 1. The lowest order Feynman diagrams for
$q\bar{q}\rightarrow \tilde{\chi}_{i}^{+}\tilde{\chi}_{j}^{-}$ and
$q\bar{q}\rightarrow \tilde{\chi}_{i}^{0}\tilde{\chi}_{j}^{0}$
processes.

FIGURE 2. The plot of cross section for $q\bar{q}\rightarrow
\tilde{\chi}_{1}^{+}\tilde{\chi}_{1}^{-}$ as a function of
$\varphi _{\mu }$ for the values of {\bf \ }$\mu =150, 300$ GeV,
M$_{2}=150, 300$ GeV and $\tan \beta =30 $ .

FIGURE 3. The plot of cross section for $q\bar{q}\rightarrow
\tilde{\chi}_{1}^{+}\tilde{\chi}_{1}^{-}$ as a function of
$\varphi _{\mu }$ and M$_{2}$, for the values of {\bf \ }$\mu=150$
GeV and $\tan \beta =30 $ .

FIGURE 4. The plot of cross section for $q\bar{q}\rightarrow
\tilde{\chi}_{1}^{+}\tilde{\chi}_{1}^{-}$ as a function of
$\varphi _{\mu }$ and $\mu$, for the values of M$_{2}=200$ GeV and
$\tan \beta =30 $ .

FIGURE 5. The plot of cross section for $q\bar{q}\rightarrow
\tilde{\chi}_{1}^{+}\tilde{\chi}_{2}^{-}$ as a function of
$\varphi _{\mu }$ for the values of {\bf \ }$\mu =150, 300$ GeV,
M$_{2}=150, 300$ GeV and $\tan \beta =30 $ .

FIGURE 6. The plot of cross section for $q\bar{q}\rightarrow
\tilde{\chi}_{1}^{+}\tilde{\chi}_{2}^{-}$ as a function of
$\varphi _{\mu }$ and M$_{2}$, for the values of $\mu =150$  GeV
and $\tan \beta =30 $ .

FIGURES 7. The plot of normal polarization for
$q\bar{q}\rightarrow \tilde{\chi}_{1}^{+}\tilde{\chi}_{2}^{-} $ as
a function of $\Theta $ for the values of $\mu =150,300$ GeV,
M$_{2}=150,300$ GeV and $\tan \beta =30$, when $\varphi _{\mu }$ =
$\pi /2$ .

 FIGURES 8. 3-dimensional plot of normal polarization for $q\bar{q}\rightarrow
\tilde{\chi}_{1}^{+}\tilde{\chi}_{2}^{-} $ as a function of
$\Theta $ and $\varphi _{\mu }$ for the values of $\mu =150$ GeV,
M$_{2}=150$ GeV and $\tan \beta =30$.

FIGURE 9. The plot of cross section for $q\bar{q}\rightarrow
\tilde{\chi}_{1}^{0}\tilde{\chi}_{1}^{0}$ as a function of
$\sqrt{s}$ for the values of $\varphi _{\mu }=0,\frac{\pi}{2},
\pi$  \ and $\frac{3\pi}{2}$ ,where M$_{1}=150$ GeV, M$_{2} = 150
$ GeV, $\mu =200$ GeV and $\tan \beta =30$.

FIGURE 10. The plot of cross section for $q\bar{q}\rightarrow
\tilde{\chi}_{1}^{0}\tilde{\chi}_{1}^{0}$ as a function of
$\sqrt{s}$ and $\mu$ , where $\varphi _{\mu } = \frac{\pi}{2}$ ,
M$_{1}=150$ GeV, M$_{2}=150$ GeV and $\tan \beta =30$.

FIGURE 11. The plot of cross section for
$q\bar{q}\rightarrow\tilde{\chi}_{1}^{0}\tilde{\chi}_{1}^{0}$
 as a function of $\varphi _{\mu }$ and $\mu$ , where\
$\sqrt{s}$ = 500 GeV, M$_{1}=150$ GeV, M$_{2}=150$ GeV and
$\tan\beta =30$.

\newpage
\begin{table}[ht]
\caption{ The cross section values for $q\bar{q}\rightarrow \tilde{\protect%
\chi}_{1}^{+}\tilde{\protect\chi}_{1}^{-}$ and $q\bar{q}\rightarrow \tilde{%
\protect\chi}_{1}^{+}\tilde{\protect\chi}_{2}^{-}$ processes for $\protect%
\varphi _{\protect\mu }=\protect\pi /3$, $\protect\mu =150,200,250$ GeV, M$%
_{2}=150,$ $200,$ $250$ GeV, and tan$\protect\beta =4,$ $10,30,50.$ }
\label{Table I.}\unitlength1mm \centering
\par
\begin{tabular}{|l|l|l|l|l|}
tan$\beta $ & M$_{2}(GeV)$ & $\mu(GeV) $ & $\sigma (q\bar{q}\rightarrow
\tilde{\chi}_{1}^{+}\tilde{\chi}_{1}^{-})$(pb) & $\sigma (q\bar{q}%
\rightarrow \tilde{\chi}_{1}^{+}\tilde{\chi}_{2}^{-})$(pb) \\ \hline
4 & 150 & 150 & 4.76 & 0.25 \\ \hline
4 & 150 & 200 & 5.41 & 0.26 \\ \hline
4 & 150 & 250 & 5.90 & 0.16 \\ \hline
4 & 200 & 150 & 4.02 & 0.15 \\ \hline
4 & 200 & 200 & 4.45 & 0.16 \\ \hline
4 & 200 & 250 & 4.93 & 0.00 \\ \hline
4 & 250 & 150 & 3.57 & 0.07 \\ \hline
4 & 250 & 200 & 3.64 & 0.00 \\ \hline
4 & 250 & 250 & 3.75 & 0.00 \\ \hline
10 & 150 & 150 & 4.76 & 0.24 \\ \hline
10 & 150 & 200 & 5.42 & 0.25 \\ \hline
10 & 150 & 250 & 5.91 & 0.15 \\ \hline
10 & 200 & 150 & 3.98 & 0.14 \\ \hline
10 & 200 & 200 & 4.42 & 0.15 \\ \hline
10 & 200 & 250 & 4.91 & 0.00 \\ \hline
10 & 250 & 150 & 3.53 & 0.06 \\ \hline
10 & 250 & 200 & 3.58 & 0.00 \\ \hline
10 & 250 & 250 & 3.66 & 0.00 \\ \hline
\end{tabular}
\par
continued
\par
\newpage
\par
\begin{tabular}{|l|l|l|l|l|}
30 & 150 & 150 & 4.75 & 0.24 \\ \hline
30 & 150 & 200 & 5.43 & 0.25 \\ \hline
30 & 150 & 250 & 5.92 & 0.15 \\ \hline
30 & 200 & 150 & 3.96 & 0.14 \\ \hline
30 & 200 & 200 & 4.40 & 0.15 \\ \hline
30 & 200 & 250 & 4.89 & 0.00 \\ \hline
30 & 250 & 150 & 3.50 & 0.06 \\ \hline
30 & 250 & 200 & 3.54 & 0.00 \\ \hline
30 & 250 & 250 & 3.60 & 0.00 \\ \hline
50 & 150 & 150 & 4.75 & 0.24 \\ \hline
50 & 150 & 200 & 5.43 & 0.25 \\ \hline
50 & 150 & 250 & 5.92 & 0.15 \\ \hline
50 & 200 & 150 & 3.95 & 0.14 \\ \hline
50 & 200 & 200 & 4.39 & 0.15 \\ \hline
50 & 200 & 250 & 4.89 & 0.00 \\ \hline
50 & 250 & 150 & 3.50 & 0.06 \\ \hline
50 & 250 & 200 & 3.53 & 0.00 \\ \hline
50 & 250 & 250 & 3.59 & 0.00 \\ \hline
\end{tabular}
\end{table}

\newpage
\begin{figure}[tbph]
\caption{{}}
 \vspace*{1.0cm}\hspace{2.5cm}
\epsfig{figure=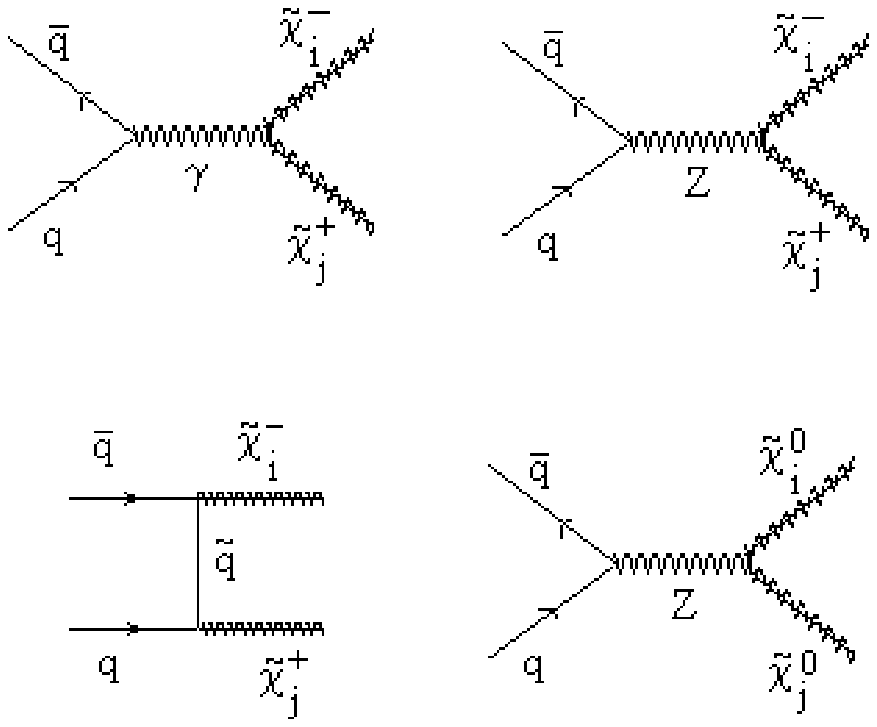,width=12cm,height=14cm,
angle=0}\vspace*{1.0cm} \label{Fig1}
\end{figure}
\newpage
\begin{figure}[tbph]
\caption{{}}
\begin{picture}(161,265)
\put(15,-255){\psfig{file=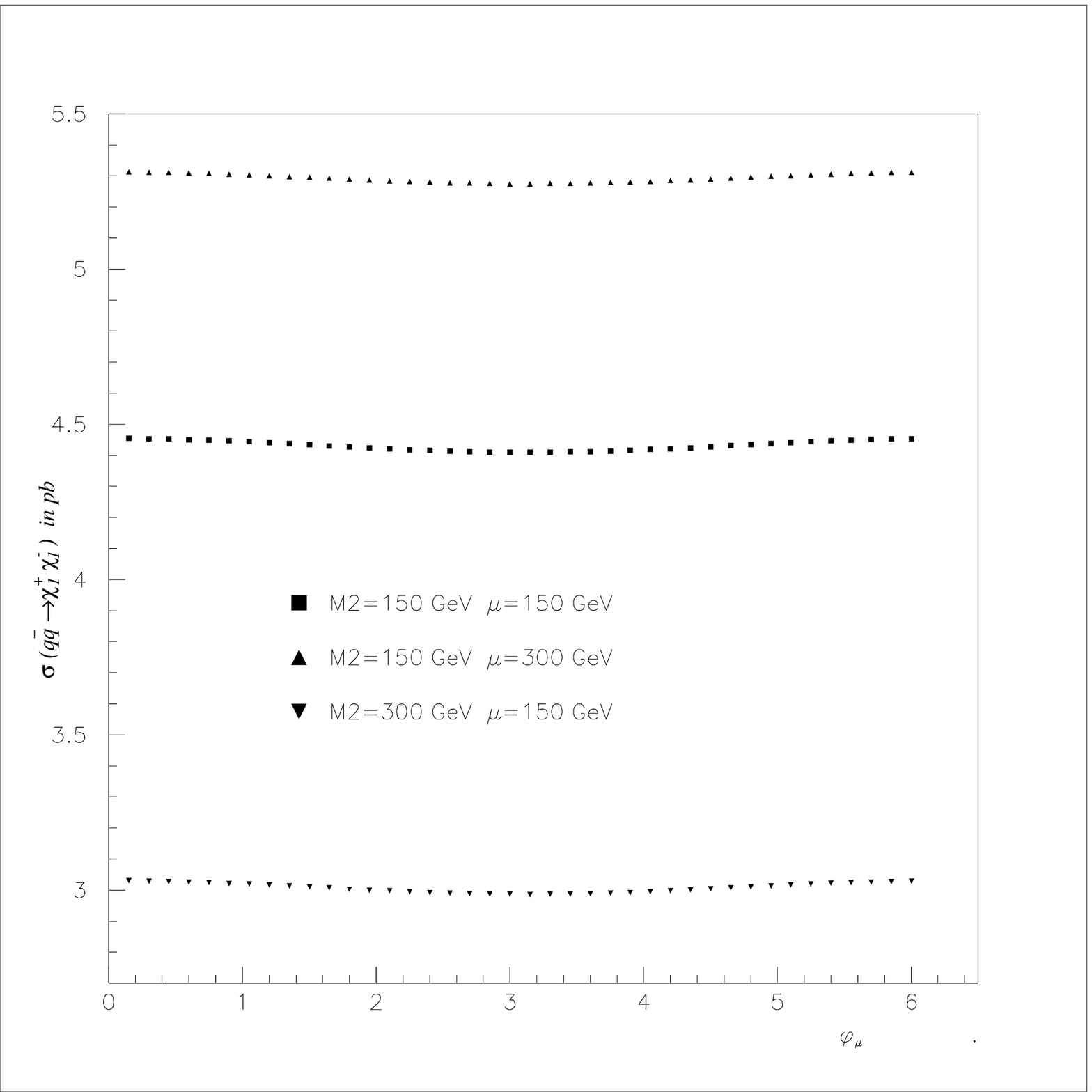,width=13cm} }
\end{picture}
\label{Fig2}
\end{figure}
\newpage
\begin{figure}[tbph]
\caption{{}}
\begin{picture}(161,265)
\put(15,-155){\psfig{file=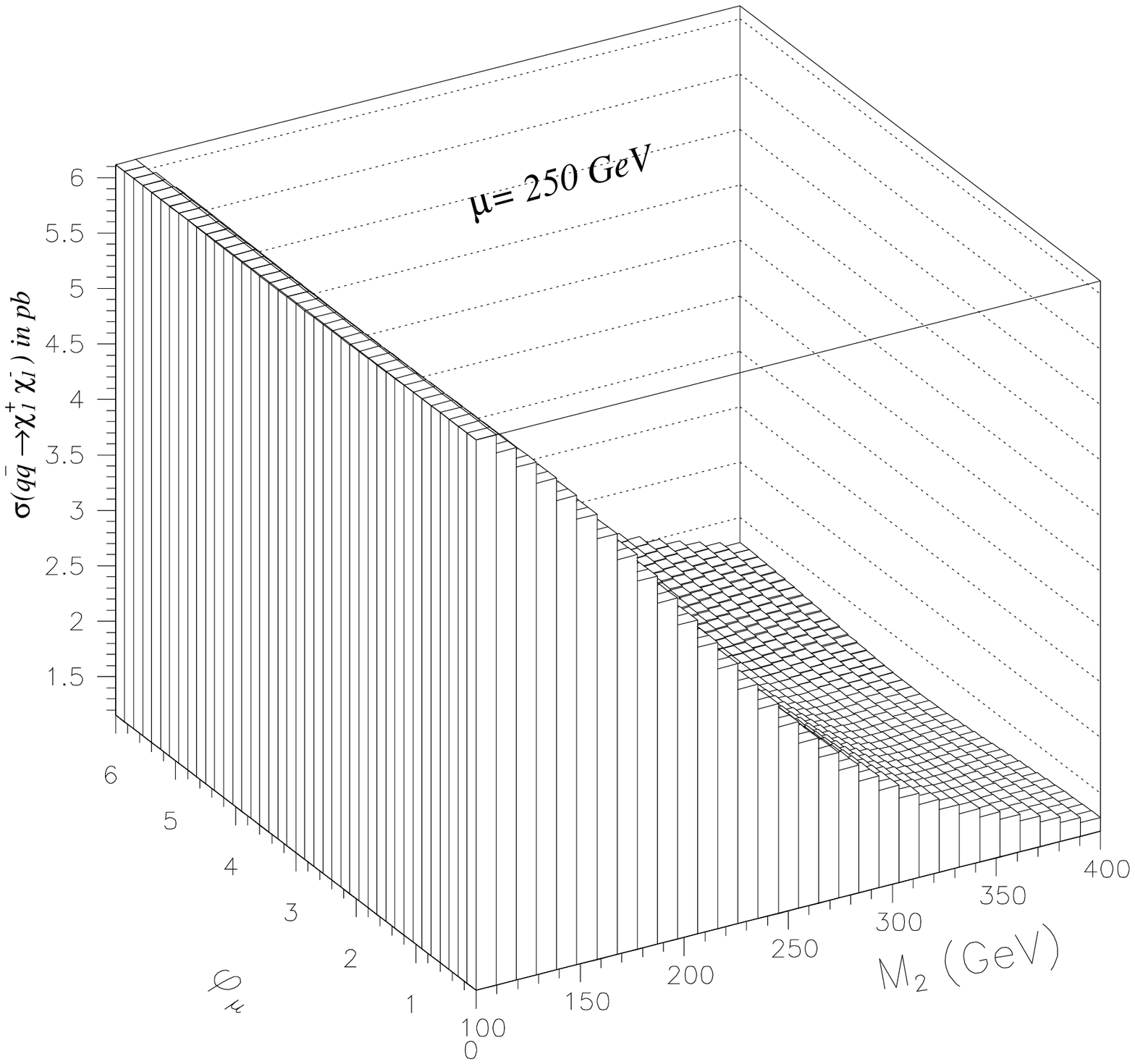,width=16cm} }
\end{picture}
\label{Fig3}
\end{figure}
\newpage
\begin{figure}[tbph]
\caption{{}}
\begin{picture}(161,265)
\put(15,-155){\psfig{file=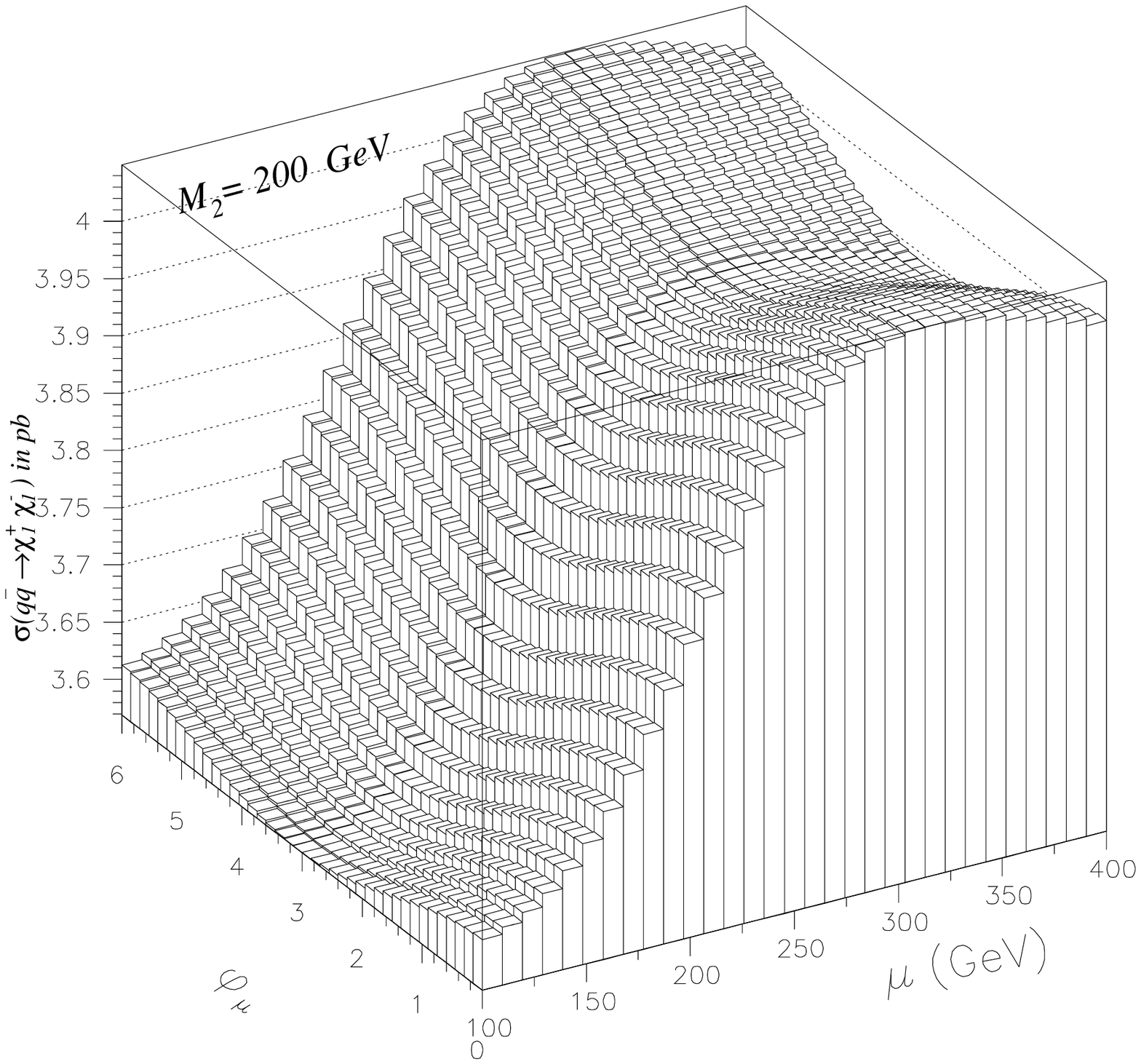,width=16cm} }
\end{picture}
\label{Fig4}
\end{figure}
\newpage
\begin{figure}[tbph]
\caption{{}}
\begin{picture}(161,265)
\put(15,-155){\psfig{file=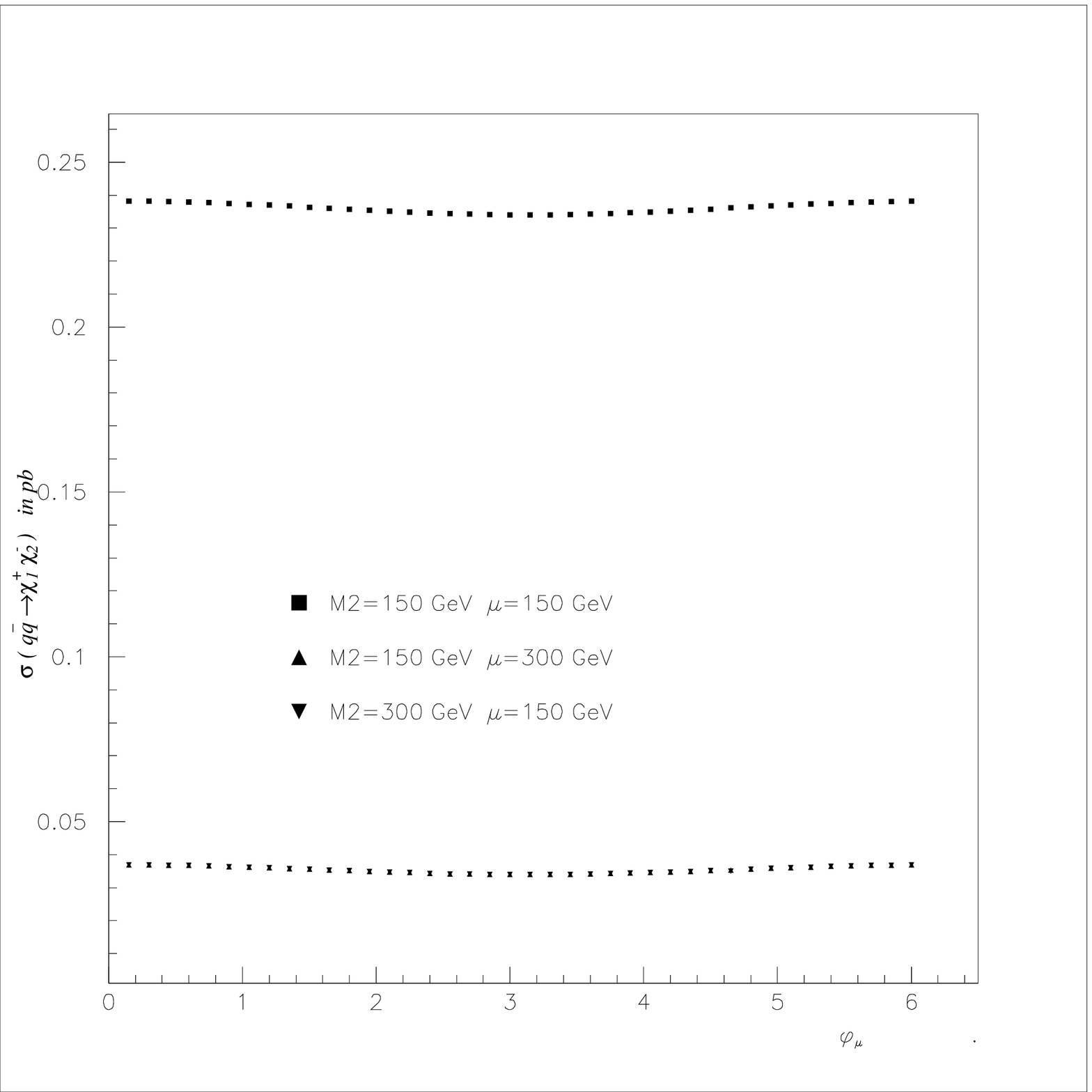,width=14cm} }
\end{picture}
\label{Fig5}
\end{figure}

\newpage
\begin{figure}[tbph]
\caption{{}}
\begin{picture}(161,265)
\put(15,-155){\psfig{file=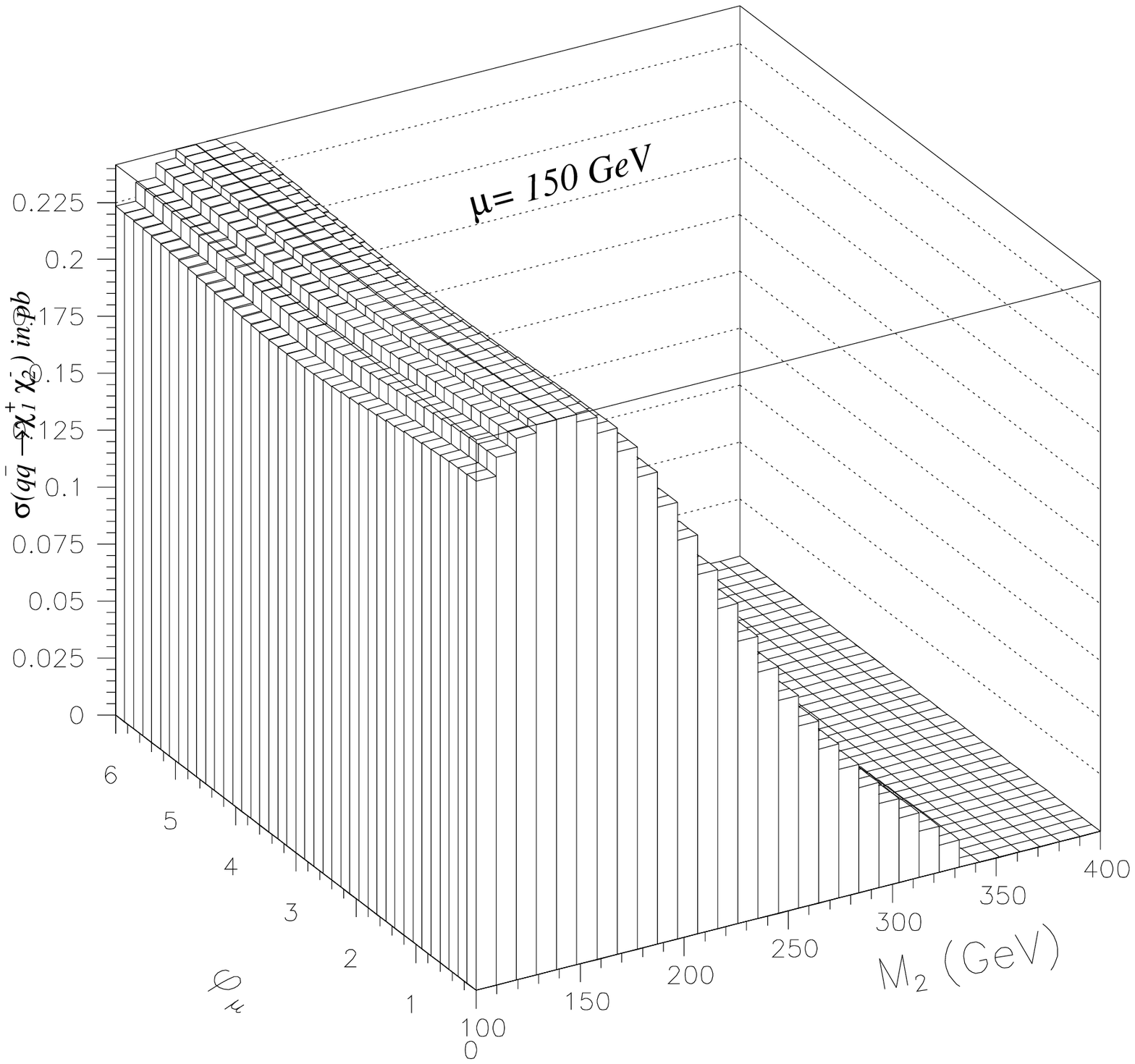,width=14cm} }
\end{picture}
\label{Fig6}
\end{figure}

\newpage
\begin{figure}[tbph]
\caption{{}}
\begin{picture}(161,265)
\put(15,-155){\psfig{file=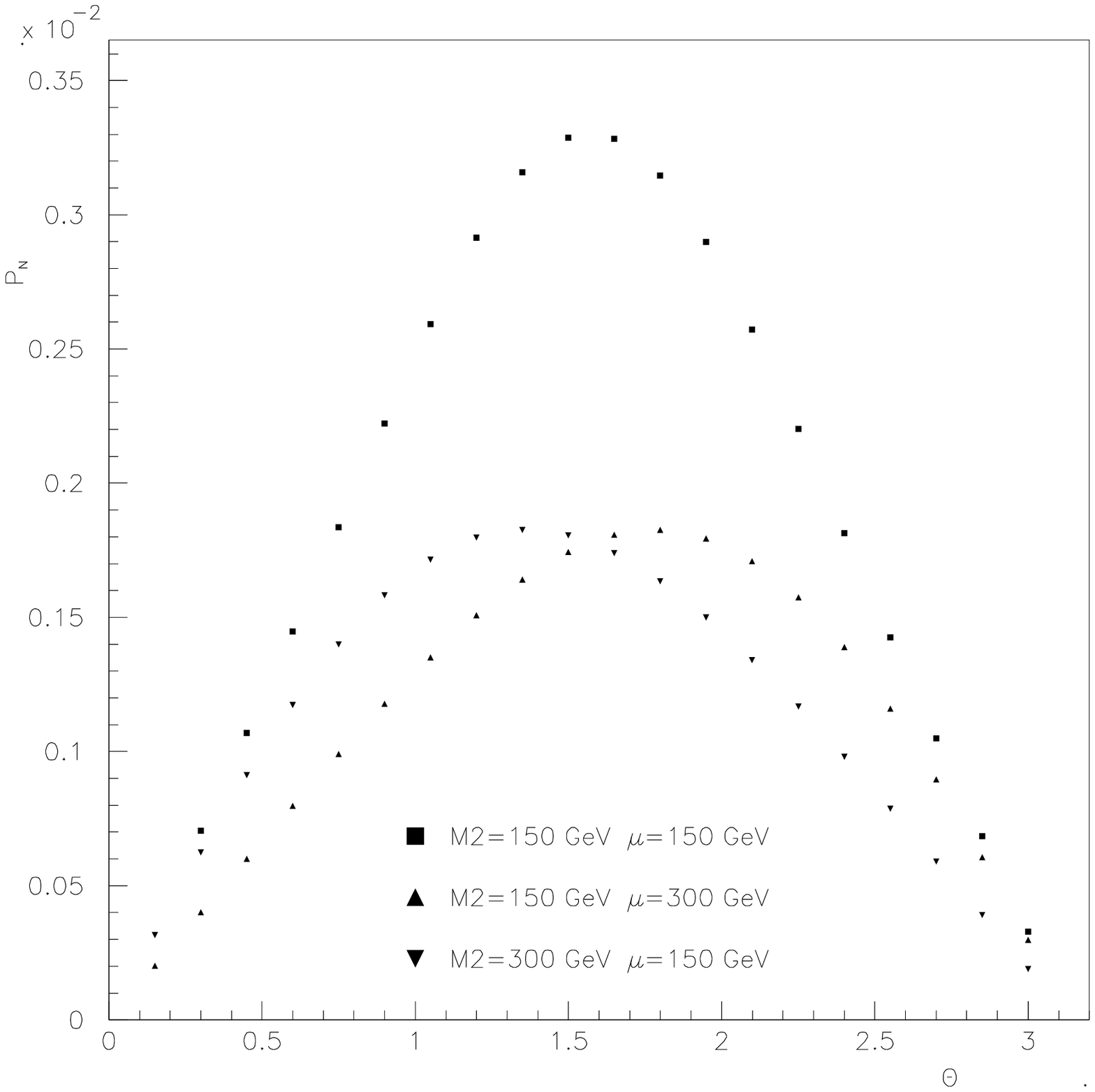,width=14cm} }
\end{picture}
\label{Fig7}
\end{figure}
\newpage
\begin{figure}[tbph]
\caption{{}}
\begin{picture}(161,265)
\put(15,-155){\psfig{file=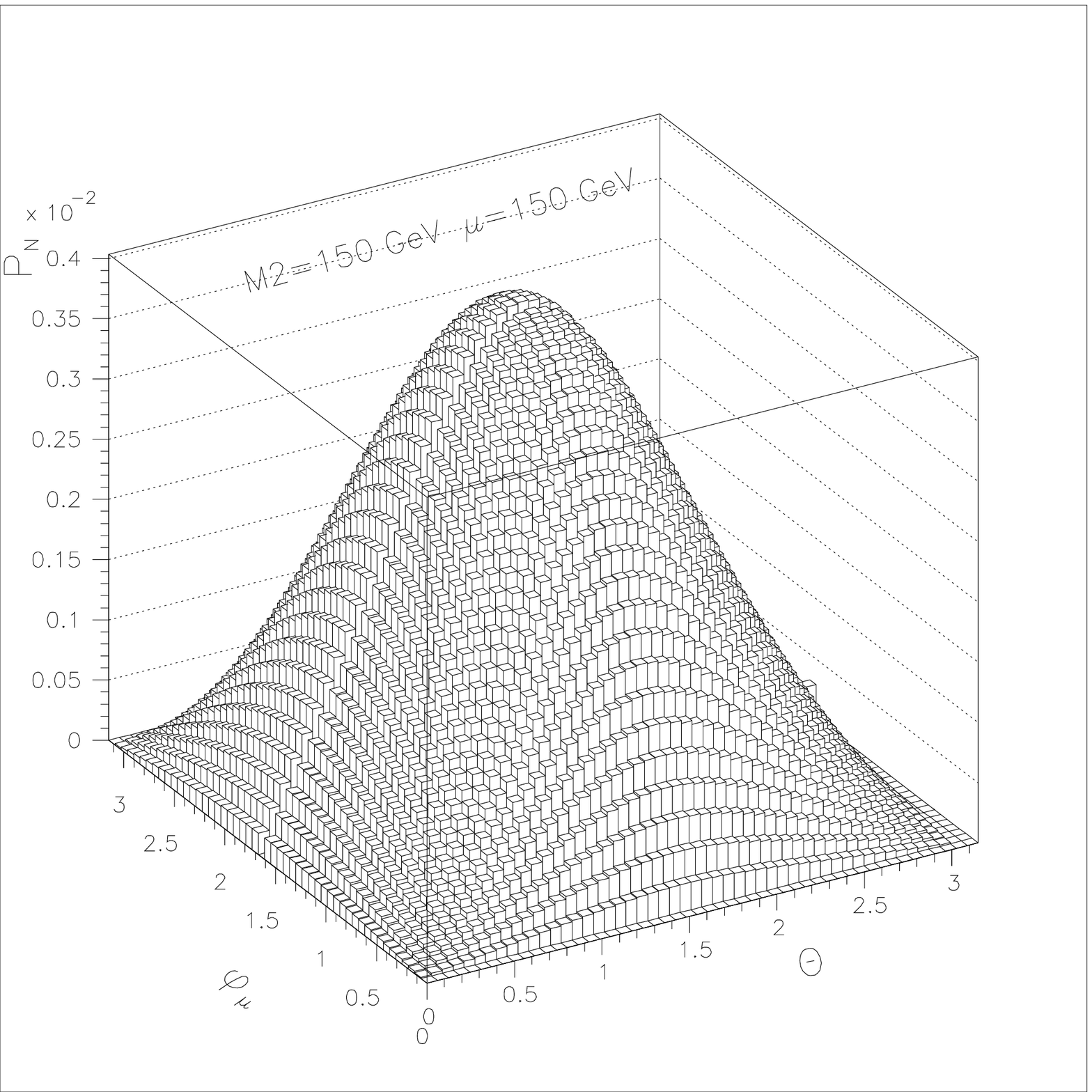,width=14cm} }
\end{picture}
\label{Fig8}
\end{figure}


\newpage
\begin{figure}[tbph]
\caption{{}}
\begin{picture}(161,265)
\put(15,-155){\psfig{file=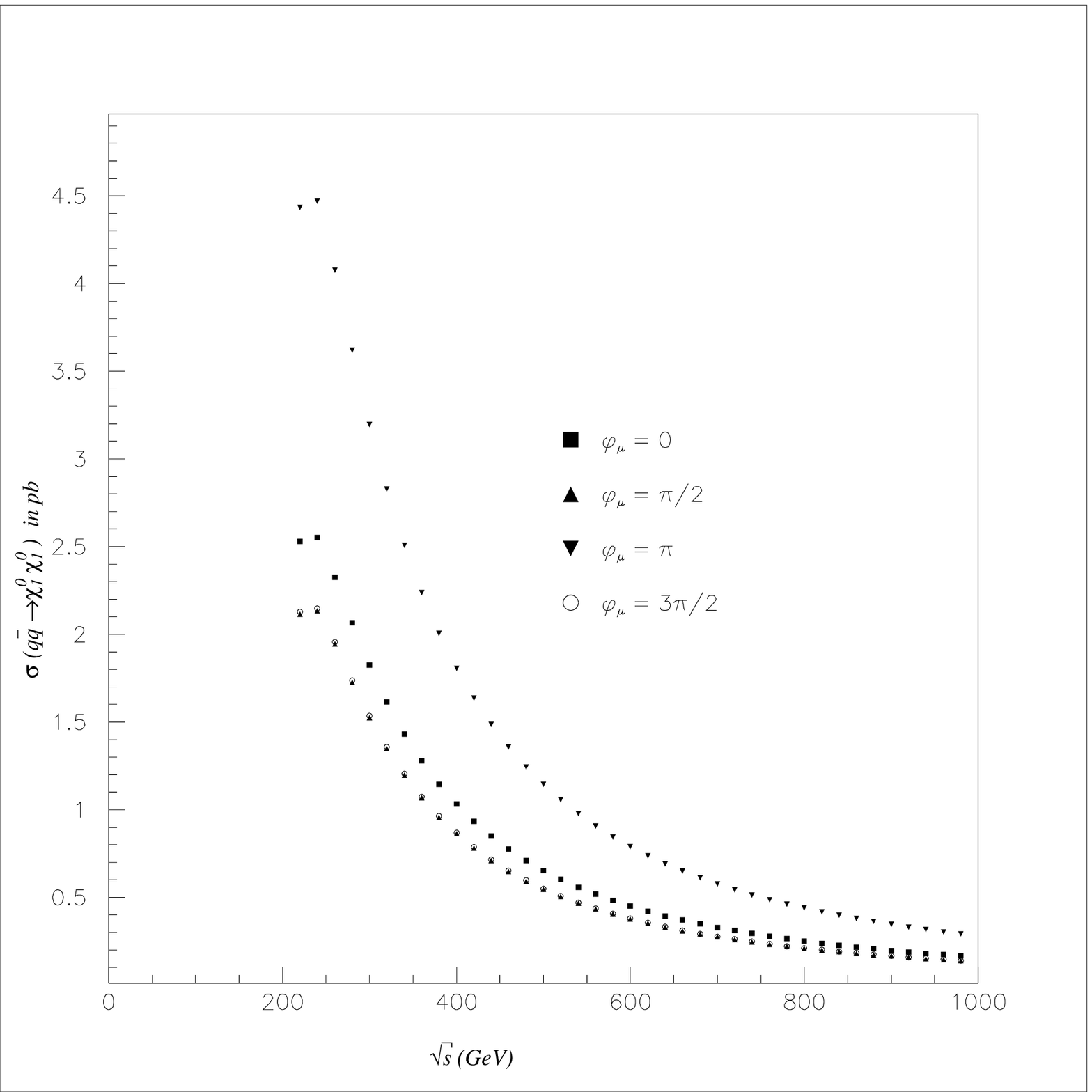,width=14cm}}
\end{picture}
\label{Fig9}
\end{figure}

\newpage

\begin{figure}[tbph]
\caption{{}}
\begin{picture}(161,265)
\put(15,-155){\psfig{file=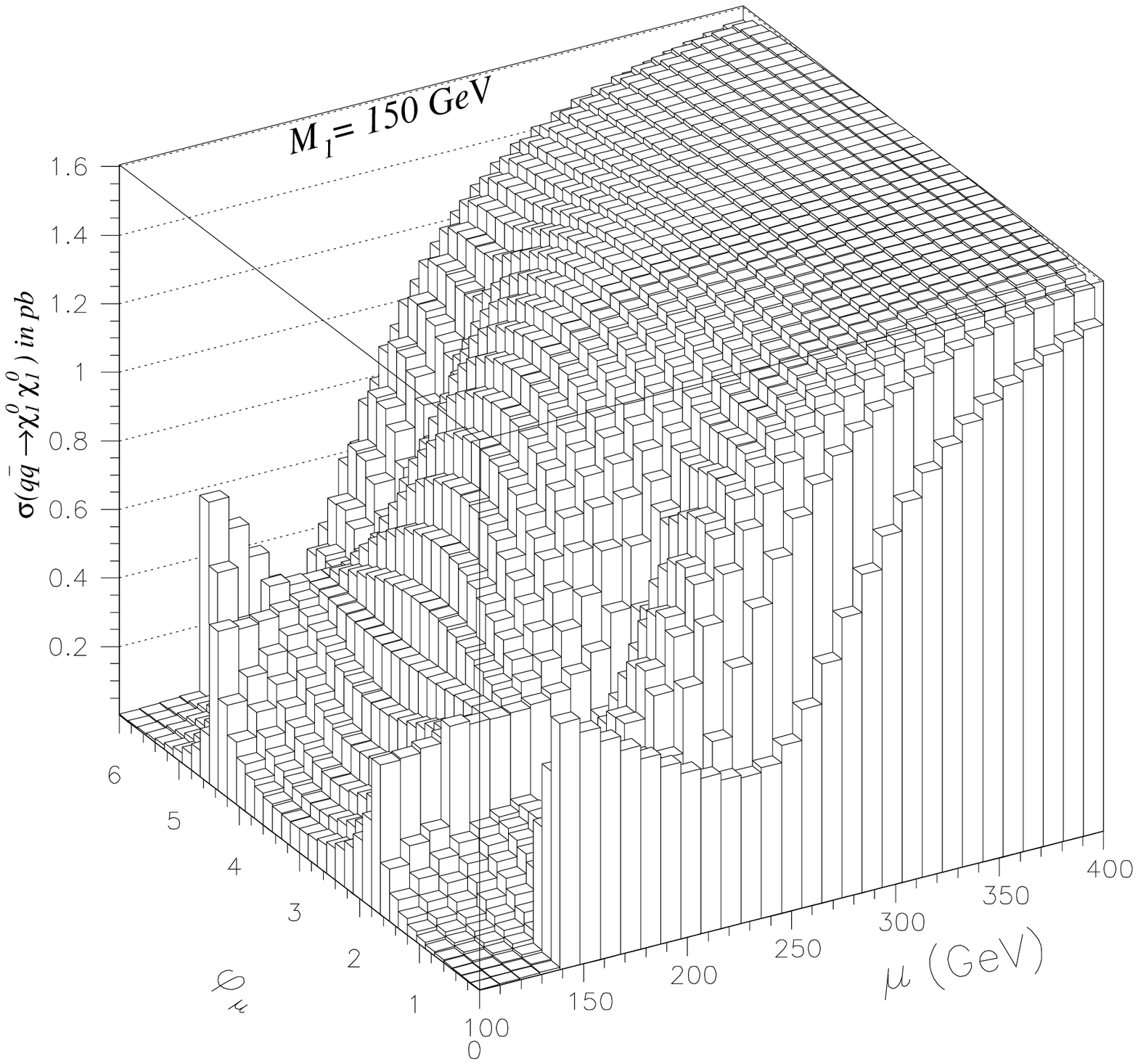,width=14cm} }
\end{picture}
\label{Fig10}
\end{figure}

\newpage
\begin{figure}[tbph]
\caption{{}}
\begin{picture}(161,265)
\put(15,-155){\psfig{file=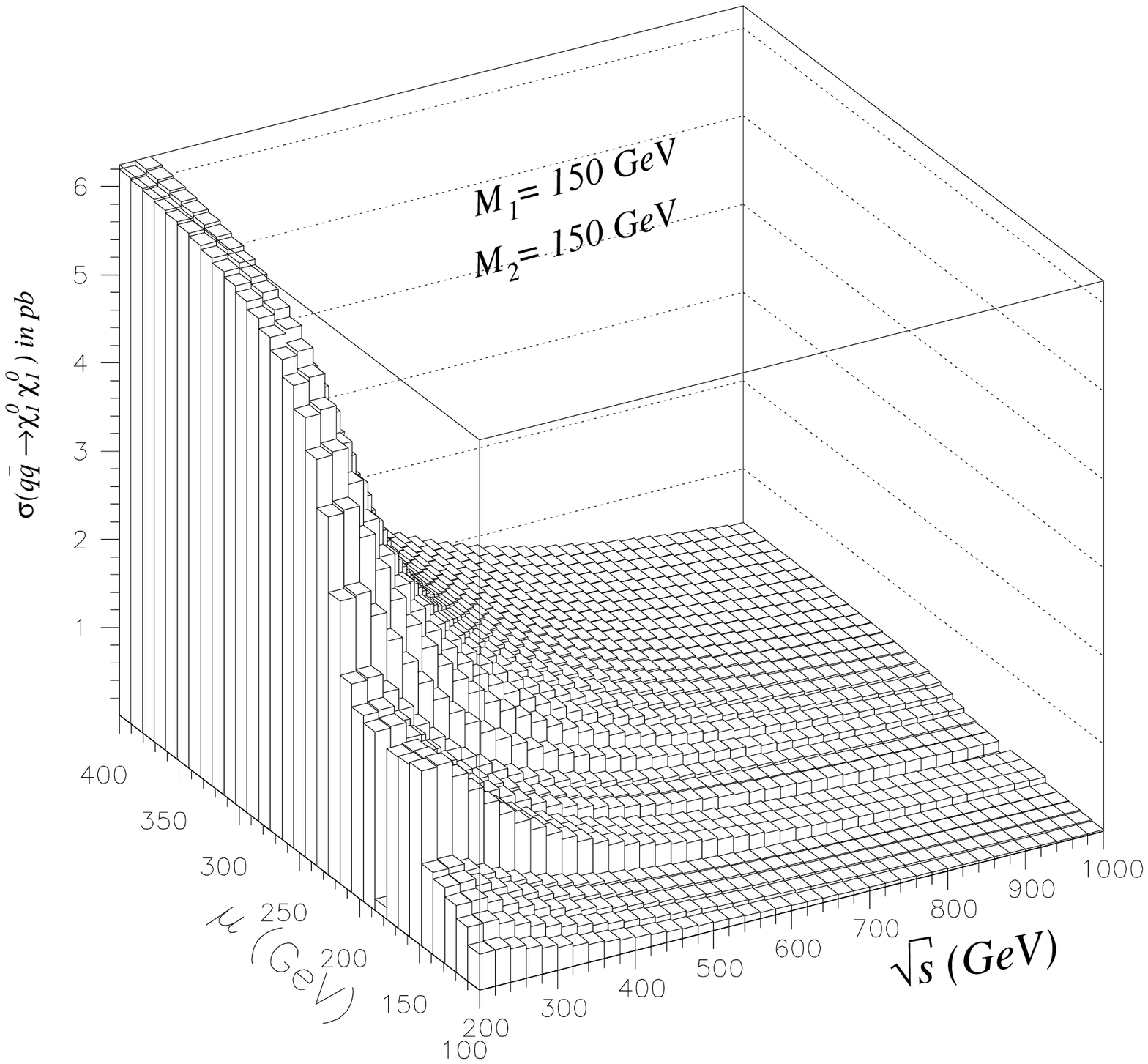,width=14cm} }
\end{picture}
\label{Fig11}
\end{figure}
\end{document}